\input{aipcheck}

\documentclass[
    ,final            
  ]
  {aipproc}

\layoutstyle{8x11single}

\usepackage{amssymb,amsmath}

\begin{document}

\newcommand{\sn}{{\rm sn}\,}
\newcommand{\cn}{{\rm cn}\,}
\newcommand{\dn}{{\rm dn}\,}

\title{Analytical solution methods for geodesic motion}


\author{E. Hackmann}{
  address={University of Bremen, ZARM, Am Fallturm, 28359 Bremen, Germany}
}

\author{C. L\"ammerzahl}{
  address={University of Bremen, ZARM, Am Fallturm, 28359 Bremen, Germany}
}


\begin{abstract}
The observation of the motion of particles and light near a gravitating object is until now the only way to explore and to measure the gravitational field. In the case of exact black hole solutions of the Einstein equations the gravitational field is characterized by a small number of parameters which can be read off from the observables related to the orbits of test particles and light rays. Here we review the state of the art of analytical solutions of geodesic equations in various space--times. In particular we consider the four dimensional black hole space--times of Pleba\'nski--Demia\'nski type as far as the geodesic equation separates, as well as solutions in higher dimensions, and also solutions with cosmic strings. The mathematical tools used are elliptic and hyperelliptic functions. We present a list of analytic solutions which can be found in the literature.
\end{abstract}

\maketitle


\section{Introduction}

In the context of General Relativity the gravitational field is the curvature of space and time. This curvature cannot be measured directly but has to be inferred from observations. As gravitational waves are not directly detected so far, the only way to measure the gravitational field is by the observation of the motion of massive objects and light moving in the gravitational field \cite{Ehlers06}. Most suitable for this task are test objects including (i) neutral point particles which are described by a normalized four velocity vector which obeys the standard geodesic equation, (ii) light rays, which are null vectors also obeying the geodesic equation, (iii) light including polarization, (iv) particles with spin obeying the Mathisson--Papapetrou equations \cite{Mathisson37,Papapetrou51}, (v) extended particles with spin and mass multipoles which are described by the Mathisson--Papapetrou--Dixon equations \cite{Dixon:1979}, (vi) electromagnetic fields which are described by the Maxwell equations minimally coupled to the gravitational field, and (vii) scalar and spinorial fields obeying the minimally coupled Klein--Gordon or Dirac equation. For many applications in astronomy it is sufficient to consider the motion of point like particles and light rays. In some cases it is necessary to take the spin and the mass multipoles of test bodies into account. Beyond the test object approach self force and radiation reaction effects have to be considered \cite{lrr-2011-7}. Here we are considering the motion of point like test bodies and light rays only. 

As the motion of test particles and light is of great importance it is most desirable to derive analytic expressions for geodesics. Beside the scientific value of analytical solutions itself, first of all analytic expressions enable a systematic study of all possible solutions and of the complete parameter space. The structure and characteristics of the solutions can be explored. This includes the derivation of observable effects like the perihelion shift or the light deflection. Furthermore, analytical solutions are a starting point for approximations in complex scenarios where analytical methods fail or for post-Newtonian approximation. They may also serve as test beds for numerical methods.


In this contribution we review analytical solution methods for the geodesic equation in various space-times beyond solutions in terms of elementary functions. We describe how solutions in terms of elliptic or hyperelliptic functions can be derived and provide a guide to the literature, which addresses such problems.

\section{Geodesic motion}

In General Relativity the motion of a point like neutral test particle is described by the geodesic equation
\begin{align}
\frac{d^2x^\mu}{ds^2} + \Gamma^\mu{}_{\nu\rho} \frac{dx^\nu}{ds} \frac{dx^\rho}{ds} = 0\label{geodesiceq}
\end{align}
where 
\begin{equation}
\Gamma^\mu{}_{\nu\rho} = \frac12 g^{\mu\sigma} \left(\partial_\nu g_{\rho\sigma} + \partial_\rho g_{\nu\sigma} - \partial_\sigma g_{\nu\rho}\right)
\end{equation}
are the Christoffel symbols, which depend on the space-time metric $(g_{\mu\nu})$, and $s$ is an affine parameter. If the metric $(g_{\mu\nu})$ is axially symmetric there are the two constants of motion
\begin{align}
E & = u_\mu \xi^\mu_{(t)}\,, \quad L_z = - u_\mu \xi^\mu_{(\varphi)} \label{genEL}
\end{align}
which are related to the Killing vectors $\xi_{(t)}$ and $\xi_{(\varphi)}$ and can be interpreted as the energy and the specific angular momentum in direction of the symmetry axes. Here $u$ denotes the four-velocity. In addition, we may normalize $g_{\mu\nu}dx^\mu dx^\nu = \epsilon$, where $\epsilon=0$ for light and $\epsilon=1$ for massive particles. However, these three constants are in general not sufficient to separate the geodesic equation \eqref{geodesiceq}. If we have even spherical symmetry we may restrict the motion to the equatorial plane but this is not possible in general. Surprisingly, a fourth constant of motion, which is not connected to an obvious symmetry, can be found for a quite wide range of space-times. In 1968 Carter \cite{Carter68} found such an additional constant of motion, which can be derived as a separation constant. As a result, the geodesic equation in a wide range of space-times can be rewritten in the form
\begin{align}
\left(\frac{dx}{d\lambda}\right)^2 & = P(x) \label{genrtheta}
\end{align}
where $x$ is (a function of) the radius $r$ or latitude $\theta$, $P(x)$ is a polynomial, and $\lambda$ is an affine parameter, which in general does not coincide with $s$. (Note that in spherically symmetric space-times often $\lambda$ can be replaced by the angle $\varphi$.) The equations for the angle $\varphi$ and the coordinate time $t$ can then be derived from \eqref{genEL} and take the form
\begin{align}
\frac{d\tilde{x}}{d\lambda} = f(r) + g(\theta)\,, \label{genphit}
\end{align}
where $\tilde{x}$ is $\varphi$ or $t$ and $f$, $g$ are rational functions. 

For example, timelike geodesic motion and even the motion of charged particles in Kerr-Newman-Taub-NUT-de Sitter space-time \cite{GriffithsPodolsky06,MankoRuiz2005} may be reduced to the form \eqref{genrtheta}, \eqref{genphit}. For lightlike motion in these space-times in addition accelerated gravitating objects may be considered \cite{GriffithsPodolsky06}. Also, timelike and lightlike geodesic motion in some higher dimensional generalizations of Kerr-Newman-de Sitter space-time may be reduced to the form \eqref{genrtheta}, \eqref{genphit} as well as Kerr and Schwarzschild space-times pierced by cosmic strings and the geodesic motion in some spherically symmetric regular black holes. A more detailed account on applications to various space-times can be found in the section on analytical solutions in the literature.

\section{Analytical solution methods}\label{sec:AS}

The analytical solution methods for differential equations of the type \eqref{genrtheta} depend on the degree of the polynomial $P$. If the degree of $P$ is two or lower than the differential equation can be solved in terms of elementary functions. If the degree is three or four the equation can be solved in terms of elliptic functions. If it is of even higher degree hyperelliptic functions are used to solve the equation. The solution methods in terms of elliptic and hyperelliptic functions are outlined in the section, where we restrict ourselves to degrees below six and only shortly mention the even more general case. 

\subsection{Elliptic functions}

Differential equations of the type
\begin{align}
\left(\frac{dx}{dy}\right)^2 & = P_{3,4}(x)\,, \quad x(y_0)=x_0 \label{genellipticdiff}
\end{align}
where $P_{3,4}$ is a polynomial of degree three or four, can be solved in terms of elliptic functions. These are meromorphic functions with two linearly independent periods $2\omega_1, 2\omega_2 \in \mathbb{C}$, i.e.~$\omega_1^{-1}\omega_2 \notin \mathbb{R}$. A non constant elliptic function has a finite number of poles in each period parallelogram $\{ z \in \mathbb{C}\,|\, \exists\, 0\leq t<1: \,z=z_0+2t\omega_1+2t\omega_2\}$, whose sum of residues has to vanish (see e.g.~\cite{Markushevich77,Hurwitz}). Historically two different approaches to elliptic function were developed named after Jacobi and Weierstrass. 

The approach by Weierstrass is based on the most simple elliptic function which has a double pole in zero with vanishing residue. This function, called the Weierstrass $\wp$ function, can be constructed as a series
\begin{align}
\wp(z;2\omega_1,2\omega_2) & = \frac{1}{z^2} + \sum_{n,m\in\mathbb{Z}, (n,m)\neq (0,0)} \left( \frac{1}{(z-2n\omega_1-2m\omega_2)^2} - \frac{1}{(2n\omega_1+2m\omega_2)^2} \right)\,,\label{defwp}
\end{align}
where the last term has to be added to achieve absolute and uniform convergence. By counting poles and using that elliptic functions without poles are constants it can be shown easily that the Weierstrass $\wp$ function solves the initial value problem (see e.g.~\cite{Markushevich77, Hurwitz})
\begin{align}
\left(\frac{dx}{dy}\right)^2 = 4x^3-g_2x-g_3\,, \quad x(0)=\infty \label{Weierstrassform}
\end{align}
where the invariants $g_2$, $g_3$ are given by $g_2 = 60 \sum_{(n,m)\neq(0,0)} (2n\omega_1+2m\omega_2)^{-4}$, $g_3 = 140 \sum_{(n,m)\neq(0,0)} (2n\omega_1+2m\omega_2)^{-6}$. Note that they are connected to the zeros $e_j$, $j=1,\ldots,3$ of $P_W(x):=4x^3-g_2x-g_3$ by
\begin{align}
e_1+e_2+e_3= 0\,, \quad -4(e_1e_2+e_1e_3+e_2e_3)=g_2\,,\quad 4e_1e_2e_3=g_3\,.
\end{align}
For the applications we have in mind here usually $P_W$ and, therefore, the zeros $e_j$ and the invariants $g_2$, $g_3$ are known instead of the periods $2\omega_{1,2}$. The periods can be calculated from the invariants by integration: in general, for two different closed integration paths $\gamma_{1,2}$,
\begin{align}\label{ellipticperiods}
\oint_{\gamma_1} \frac{dz}{\sqrt{P_W(z)}} \neq \oint_{\gamma_2} \frac{dz}{\sqrt{P_W(z)}}
\end{align}
as the zeros $e_j$ are singularities of the integrand. All periods $(2n\omega_1+2m\omega_2)$, $n,m\in\mathbb{Z}$, can then be determined by integrating along all possible paths. The basic periods $2\omega_{1,2}$ can e.g.~be determined by
\begin{equation}
\begin{aligned}
\omega_1 & = \int_{e_1}^{e_2} \frac{dz}{\sqrt{P_W(z)}} = \int_{e_3}^{\infty} \frac{dz}{\sqrt{P_W(z)}} \,, \\
\omega_2 & = \int_{e_2}^{e_3} \frac{dz}{\sqrt{P_W(z)}} = \int_{-\infty}^{e_1} \frac{dz}{\sqrt{P_W(z)}}
\end{aligned}
\end{equation}
if all $e_j$ are real and ordered as $e_1<e_2<e_3$ or if $e_1=\bar{e}_2$ and $e_3 \in \mathbb{R}$, where the paths are assumed not to cross each other. (Here we used the convention that $\omega_1\in \mathbb{R}$; other conventions use $\omega_1=\bar{\omega}_2$ if $e_1=\bar{e}_2$.)

The series expansion \eqref{defwp} is not very well suited to actually compute the Weierstrass $\wp$ function as it converges quite slowly. Alternatively it can be calculated iteratively \cite{Coquereauxetal1990}, using a power series with Eisenstein series as coefficients, or in terms of theta functions \cite{Markushevich77, Hurwitz}. The Riemann theta function is defined as
\begin{align}
\theta[\tau v+w](z;\tau) & = \sum_{m \in \mathbb{Z}^g} \exp( \pi i (m+v)^t (\tau (m+v) + 2z + 2w) )\,, \label{deftheta}
\end{align}
where $z\in\mathbb{C}^g$, $\tau$ is a $g\times g$ symmetric matrix with positive definite imaginary part, $\tau v+w \in \mathbb{C}^g$ is the characteristic, and $g$ is the genus, here $g=1$.

The general differential \eqref{genellipticdiff} can be solved in terms of the Weierstrass elliptic function by converting the problem to the standard form \eqref{Weierstrassform}. If $P_{3,4}$ is of degree four first apply the substitution $x=\xi^{-1}+x_P$ where $P_{3,4}(x_P)=0$ to convert to a polynomial of degree three. Subsequently, or if $P_{3,4}$ was of degree three in the first place, substitute $\xi=\frac{1}{b_3}(4z+\frac{b_2}{3})$ if $P=\sum_i b_ix^i$ is the polynomial of degree three. Then 
\begin{equation}\label{general_g2g3}
\begin{aligned}
g_2 & = \frac{1}{16} \left( \frac{4}{3} b_{2}^2 - 4 b_{1} b_{3} \right) \,, \\
g_3 & = \frac{1}{16} \left( \frac{1}{3} b_{1} b_{2} b_{3} - \frac{2}{27} b_{2}^3 - b_{0} b_3^2 \right) \,,
\end{aligned}
\end{equation}
and $z=\wp(y-y_{\rm in})$ where $y_{\rm in}=y_0+\int_{z_0}^{\infty} \frac{dz}{\sqrt{P_W(z)}}$ depends only on the initial values.

The approach by Jacobi is based on the inversion of elliptic integrals. The sine amplitude function $\sn$ can be defined as
\begin{align}
z = \int_0^w \frac{dt}{\sqrt{(1-t^2)(1-k^2t^2)}} =: F(w;k) \quad \Rightarrow \quad \sn(z;k)=w
\end{align}
where $0\leq k\leq 1$ is the modulus, $w\in [0,1]$, and $z \in \mathbb{R}$. It has poles in $iK(k')$ and $2K(k)+iK(k')$ where $K(k)=F(1;k)$ is the complete elliptic integral of the first kind and $(k')^2=1-k^2$. The sine amplitude $\sn$ can also be written in terms of the Riemann theta function. Indeed, it is often defined in this way,  
\begin{align}
\sn(z;k) = k^{-\frac{1}{2}} \frac{\theta[\frac{1}{2}](u;\tau)}{\theta[\frac{\tau+1}{2}](u;\tau)}\,, 
\end{align}
where $\tau=\frac{iK(K')}{K(k)}$ and $u=\frac{z}{2K(k)}$.

The general differential equation \eqref{genellipticdiff} can be solved in terms of Jacobian elliptic functions by applying a substitution which converts the problem to the form
\begin{align}
\left(\frac{d\tilde{x}}{dy}\right)^2 & = (1-\tilde{x}^2)(1-k^2\tilde{x}^2)\,.
\end{align}
This substitution depends on the degree of $P_{3,4}$ and its number of complex zeros as well as on the type of orbit you want to obtain. For a list see e.g.~\cite{Abramowitz}.

\subsection{Solutions in terms of hyperelliptic functions}

In the previous section we mentioned that both the Weierstrass and the Jacobi elliptic functions can be defined in terms of theta functions. To generalize the results on elliptic functions to differential equations of the form
\begin{align}
\left(\frac{dx}{dy}\right)^2 & = P_{5,6}(x)\,, \quad x(y_0)=x_0\label{genhyperdiff}
\end{align}
where $P_{5,6}$ is of degree five or six, we first need to define more general functions. In \eqref{ellipticperiods} we found that all periods of the solution function of \eqref{genellipticdiff} can be determined by different integration paths. As there the integrand had four singular points we had two independent periods. Here, the corresponding integrand has six singular points and, therefore, we get four independent periods along four integration paths $a_{1,2}$, $b_{1,2}$, which may be chosen such that integrals over $a_{1,2}$ are real. However, a function on $\mathbb{C}$ cannot have more than two independent periods. This means that we have to consider more general functions. Let us first introduce the holomorphic differentials $dz_i$ and the meromorphic differential $dr_i$ by
\begin{align}
dz_i & := \frac{t^{i-1} dt}{\sqrt{P(t)}} \label{holomorphdiff} \,,\\ 
dr_i & := \sum_{k=i}^{2g+1-i} (k+1-i) a_{k+1+i} \frac{t^k dt}{4\sqrt{P(t)}} \, , \label{meromorphdiff}
\end{align}
where $P(t)=4t^{2g+1}+\sum_{n=0}^{2g} a_nt^n$ is a polynomial of degree $2g+1$ and $g$ is the genus. We may then define the corresponding $g\times g$ period matrices of the first kind $(2\omega_1)$, $(2\omega_2)$ and of the second kind $(2\eta_1)$, $(2\eta_2)$ by 
\begin{equation}\label{periodmatrices}
\begin{aligned}
(2 \omega_1)_{ij} &:= \oint_{a_j} dz_i\,, & \qquad (2 \omega_2)_{ij} &:= \oint_{b_j} dz_i\,, \\
(2 \eta_1)_{ij} &:= - \oint_{a_j} dr_i\,, & \qquad (2 \eta_2)_{ij} &:= - \oint_{b_j} dr_i\, .
\end{aligned}
\end{equation}
Here $a_1,\ldots,a_g$, $b_1,\ldots,b_g$ are the $2g$ independent closed integration paths. If $P$ has only real zeros $e_j$ ordered as $e_j<e_{j+1}$ one may e.g.~choose $a_j$ as the path from $e_{2j-1}$ along the real axes to $e_{2j}$ and back (with reversed sign of the square root) and the path $b_j$ from $e_{2j}$ along the real axes to $e_{2j+1}$ and back (again with reversed sign of the square root).

To generalize the solution methods outlined in the previous section we first need to consider the Jacobi inversion problem 
\begin{align}
y_i = \sum_{j=1}^g \int_{\infty}^{x_j} \frac{t^{i-1}dt}{\sqrt{P(t)}}\,, \quad i=1,\ldots,g\,. \label{Jacobiinversion}
\end{align}
Note that for $g=1$ and $a_2=0$ we recover \eqref{Weierstrassform}. The $g$ solutions $x_j(y_1,\ldots,y_g)$ of \eqref{Jacobiinversion} can be given in terms of generalized Weierstrass functions. These are defined by the theta function via the Kleinian sigma function
\begin{align}
\sigma(z;\omega_1,\omega_2) & = C e^{iz^t\kappa z} \theta[K_\infty](z;\omega_1^{-1}\omega_2)\,,\\
\wp_{ij}(z;\omega_1,\omega_2) & = - \frac{\partial}{\partial z_i} \frac{\partial}{\partial z_j} \log \sigma(z;\omega_1,\omega_2)\,,
\end{align}
where $z\in \mathbb{C}^g$, $\kappa = \eta(2\omega_1)^{-1}$, and $C$ is a constant with does not matter in the present context. The vector of Riemann constants with base point at infinity, $K_\infty \in \mathbb{C}^g$ can be calculated by
\begin{align}
K_{\infty,i}=(2\omega_1)^{-1} \sum_{j=1}^g \int_{\infty}^{e_{2j}} \frac{t^{i-1}dt}{\sqrt{P(t)}}
\end{align}
where $e_i$ are the zeros of the polynomial $P$. This means that $K_\infty$ is a sum of normalized half periods, in particular $K_\infty=\tau(\frac{1}{2},\frac{1}{2})^t + (0,\frac{1}{2})^t$ if $g=2$. Note that $\tau=\omega_1^{-1}\omega_2$ is always symmetric with positive definite imaginary part. For further details see e.g.~\cite{BuchstaberEnolskiiLeykin97}. The solutions of \eqref{Jacobiinversion} are then given by the solutions of
\begin{align}
x^g + \sum_{i=1}^g \wp_{gi}(y_1,\ldots,y_g) x^{i-1} = 0\,.
\end{align}

Let us turn back to the differential equation \eqref{genhyperdiff} and genus $g=2$. If $P_{5,6}$ is of degree six the equation can be reformulated as $\tilde{x} \frac{d\tilde{x}}{dy}=\sqrt{P_5(\tilde{x})}$ by a substitution $x=\tilde{x}^{-1}+x_P$, where $x_P$ is a zero of $P_{5,6}$ and $P_5$ is a polynomial of degree five. Such a differential equation, or, if $P_{5,6}$ was of degree five in the first place, can then be rewritten in the form 
\begin{align}
t^{i-1}\frac{dt}{dy}=\sqrt{P(t)}\,, \quad t(y_0)=t_0\,, \label{normhyperdiff}
\end{align}
where $i=1$ or $i=2$, by an appropriate normalization. We may then consider this equation as part of the Jacobi inversion problem,
\begin{align}
\int_{\infty}^t \frac{\tilde{t}^{i-1} d\tilde{t}}{\sqrt{P(\tilde{t})}} = y - y_0 - \int_{t_0}^\infty \frac{\tilde{t}^{i-1} d\tilde{t}}{\sqrt{P(\tilde{t})}} =: y_i\,.\label{partJacobi}
\end{align}
We may then find the solution to \eqref{normhyperdiff} as the limiting case $x_2\to\infty$ of the Jacobi inversion problem \eqref{Jacobiinversion} in the following way: first observe that 
\begin{align}
t & := x_1 = \lim_{x_2\to\infty} \frac{x_1x_2}{x_1+x_2} = \lim_{x_2\to\infty} \frac{\wp_{12}(y_1,y_2)}{\wp_{22}(y_1,y_2)}\nonumber\\
& = \lim_{x_2\to\infty} \frac{\sigma\sigma_{12}-\sigma_1\sigma_2}{\sigma_2^2-\sigma\sigma_{22}} (y_1,y_2)\,,\label{solx}
\end{align}
where $\sigma_i(z)$ is the derivative of $\sigma$ with respect to $z_i$. From the comparison of \eqref{Jacobiinversion}, where $g=2$ and $x_2\to\infty$, with \eqref{partJacobi} we may identify $y_i$ in \eqref{solx} with our physical coordinate $y_i$ in \eqref{partJacobi}. Fortunately, we get automatically rid of the other $y_j$, $j\neq i$, by the same limiting process $x_2\to\infty$. This is because the set of zeros of the theta function $z\mapsto \theta[K_\infty]((2\omega)^{-1} z)$, which is a one dimensional submanifold of $\mathbb{C}^2$, is given by all vectors $z=(z_1,z_2)$ which can be written as $z_i=\int_\infty^x dz_i$ with the same $x$ and $dz_i$ as in \eqref{holomorphdiff}, see e.g.~\cite{Mumford83}. This is exactly true for the vector $(y_1,y_2)$, which means that we may write $y_j=f(y_i)$ for some function $f$. As the zeros of the theta function are also zeros of $\sigma$ we can simplify \eqref{solx} to
\begin{align}
t & = - \frac{\sigma_1}{\sigma_2} (y_1,f(y_1)) \quad \text{ or } \quad t = - \frac{\sigma_1}{\sigma_2} (f(y_2),y_2)\,.
\end{align}
with $y_i$ as defined in \eqref{partJacobi}.

Note that the solution method for differential equations of the type \eqref{genhyperdiff} can be generalized even further. As long as we have on the right hand side of \eqref{genhyperdiff} a polynomial of any degree, we may consider the equation as part of Jacobi's inversion problem. As an additional complication we have to find $g-1$ condition on the $y_i$'s on left hand side in \eqref{Jacobiinversion} to restrict the problem to a one dimensional submanifold of $\mathbb{C}^g$. For details see \cite{EnolskietalJGP11}.

\section{Analytical solutions of geodesic equation in various space-times}\label{sec:literature}

In this section we will collect applications of the methods outlined in the section on geodesic motion to various space-times. For older literature we refer to Sharp \cite{Sharp1979} who collected most of the papers on geodesic motion in Kerr-Newman space-time and subclasses, which were available at that time. Partly this is still quite complete but we also try to update his collection (with respect to analytical solutions). Note that we only consider analytical solutions to general timelike and lightlike geodesics (with an electric or magnetic charge, as applicable). In particular, we do not list the vast literature on equatorial motion in Kerr space-time. Of course, we do not claim that our list is complete.

\subsection{Particle motion in standard black hole space-times}

\subsubsection{Schwarzschild} 

Regarding analytical solution methods the list of Sharp already contained the complete set of solutions. Most notably, this includes the works by Hagihara \cite{Hagihara}, who derived the analytical solutions in terms of Weierstrass elliptic functions, and Darwin \cite{Darwin1959,Darwin1961}, who used Jacobian elliptic functions and integrals. 

Accordingly, observables like the perihelion shift, the scattering angle of flyby orbits, the light deflection (which is the flyby orbit of light) can be described in terms of (complete) elliptic integrals. 

\subsubsection{Reissner-Nordstr\"om} 

Surprisingly, the analytical solutions to the geodesic equation in Reissner-Nordstr\"om space-time seem to be considered first only in 1983 by Gackstatter \cite{Gackstatter1983} although it can be handled completely analogously to the Schwarzschild case. He studied bound timelike geodesics and light in terms of Jacobian elliptic integrals and functions. Recently, Slez\'{a}kov\'{a} \cite{Slezakova2006} gave a comprehensive analysis of arbitrary timelike, lightlike, and even spacelike geodesics. Grunau and Kagramanova \cite{Grunau2011} solved the equations of motion of electrically and magnetically charged particles in Reissner-Nordstr\"om space-time in terms of Weierstrass elliptic functions. Gackstatter and Grunau/Kagramanova also considered the perihelion shift and the deflection of light, which can be given in terms of (complete) elliptic integrals.

\subsubsection{Taub-NUT} 

Timelike geodesics were studied by Kagramanova et al \cite{Kagramanova2010} in terms of Weierstrass elliptic functions. One particular feature of the geodesics in NUT space-times is that the orbits do not lie on an orbital plane but, instead, on an orbital cone with an opening angle which depends on the NUT parameter, which can also be considered as a gravitomagnetic mass. In particular, the issue of geodesic incompleteness at the horizons as discussed by Misner and Taub \cite{MisnerTaub69}, and the analytic extension of Miller, Kruskal and Godfrey \cite{MillerKruskalGoodfrey71} was addressed.

\subsubsection{Kerr} 

Most of the older literature on Kerr space-time is concerned with the much simpler particular case of equatorial geodesics. We refer to Sharp \cite{Sharp1979} here for these early works. Note that in terms of the proper time (or the corresponding affine parameter for light) the equations of motion are still coupled. Therefore, most of the analytical solutions before the introduction of the Mino time \cite{Mino03} implicitly included integrals over the latitude or the radius, see e.g. Kraniotis \cite{Kraniotis2005} or Slez\'{a}kov\'{a} \cite{Slezakova2006} for a review. As notable exception, \v{C}ade\v{z} et al \cite{Cadezetal1998} introduced already in 1998 a similar parameter (called $P$, see their equation (34)) as they considered the motion of light. 

After the introduction of the Mino time, in 2009 Fujita and Hikida \cite{FujitaHikida09} used this new affine parameter to derive the analytical solution for bound timelike geodesics in terms of Jacobian elliptic functions. They also elaborated the expressions for perihelion shift and Lense-Thirring effect in terms of Jacobian elliptic integrals. General timelike geodesics and lightlike motion were treated shortly after that by Hackmann \cite{HackmannDiss} in 2010. Recently, also an analytical expression for the gravitomagnetic clock effect was derived \cite{Hackmannetal2013b}. 

Note that Kraniotis \cite{Kraniotis2011} also derived analytical solutions for lightlike geodesics in terms of hypergeometric functions.

\subsubsection{Kerr-Newman} 

Charged particle motion was considered by Hackmann and Xu \cite{Hackmannetal2013} in terms of Weierstrass elliptic functions. Note that also a magnetic charge of the black hole was included. Due to many parameters involved the number of orbit configurations is very large. 
This includes orbits crossing the horizons or $r = 0$. For the sake of completeness, also black holes endowed with magnetic charge has been considered. This has a big impact on the orbital types: Not only
does the motion deviate from the symmetry to the equatorial plane but also stable off-equatorial circular orbits outside the horizon do exist in this case, which are not possible elsewhere. 

\subsubsection{Schwarzschild-de Sitter}

This space-time is also named after Kottler. Note that on the level of the differential equation, lightlike geodesics in Schwarzschild-de Sitter space-time are identical with the lightlike equations of motion for Schwarzschild, as the cosmological constant can be absorbed in the definition of just a single parameter. Analytical solution are given e.g. in Gibbons et al \cite{Gibbonsetal2008}. General timelike geodesics in Kottler space-time can be treated in terms of hyperelliptic functions as elaborated by Hackmann and L\"ammerzahl \cite{HackmannLaemmerzahl2008,HackmannLaemmerzahl08b}. 

The motivation for searching for an analytical solution of the geodesic equation in Schwarzschild-de Sitter space-time was to investigate the suggestion being made that the Pioneer anomaly could originate from the influence of the cosmological constant on the motion of test bodies. With the above analytic solution it could be shown that the effect of a cosmological constant is of the order of cm only and, thus, cannot be responsible for the the anomalous acceleration which has been unambiguously measured. In fact, later the anisotropic thermal radiation could be identified as origin of the anomalous acceleration \cite{RieversLaemmerzahl11}. 

\subsubsection{Reissner-Nordstr\"om-de Sitter} 

The equations of motion for general timelike geodesics were solved in \cite{Hackmannetal2008}. The motion of photons was very recently analytically calculated by Villanueva et al \cite{Villanuevaetal2013} for a negative cosmological constants using Weierstrass elliptic functions.

\subsubsection{Taub-NUT-de Sitter} 

The transition from Taub-NUT to Taub-NUT-de Sitter is similar to the transition of Schwarzschild to Schwarzschild-de Sitter. For a not too big cosmological constant an additional cosmological horizon appears which enlerges the possible orbital types in an obvious way. The timelike motion in this case has been analyzed in \cite{Hackmannetal2009}.

\subsubsection{Kerr-de Sitter} 

The equations of motions for timelike geodesics were analytically solved by Hackmann et al \cite{Hackmannetal2009,Hackmannetal2010} in terms of hyperelliptic functions. Note that Kraniotis \cite{Kraniotis2011} also derived analytical solutions for lightlike geodesics in terms of hypergeometric functions.

\subsubsection{Kerr-Newman-Taub-NUT-de Sitter} 

The general solution for timelike geodesic was shortly outlined in \cite{Hackmannetal2009}. Analytical expressions for the perihelion shift and the Lense-Thirring effect were derived in \cite{HackmannLaemmerzahlPRD12}.

\subsection{Particle motion in higher dimensional space-times}

\subsubsection{Spherical symmetry}
Timelike and lightlike motion in higher dimensional Schwarzschild, Reissner-Nordstr\"om, Schwarzschild-de Sitter, and Reissner-Nordstr\"om-de Sitter was considered in \cite{Hackmannetal2008} and more generally in \cite{EnolskietalJGP11}.

\subsubsection{Myers-Perry}

Myers-Perry space-times are higher dimensional generalizations of the 4-dimensional Kerr space-time. Kagramanova and Reimers \cite{KagramanovaReimers2012} derived analytical solution to the geodesic equation in a five dimensional Myers-Perry space-time assuming that the two parameters of rotation coincide. Their results are valid for both massive test particles and light. Also the various observables, that is, the two types of perihelion shift as well as two types of Lense-Thirring effect have been derived. 

\subsubsection{Black rings}
In higher dimensions black holes may have a more complicated horizon topology than in four dimensions. In five dimensions, the horizon may have the topology of a ring which are accordingly called black ring space-times. Grunau et al \cite{Grunauetal2012} considered geodesics in a singly spinning black ring spacetime in five dimensions. In general the Hamilton-Jacobi equation is not separable for these space-times. For the special cases where it is separable Grunau et al derived analytical solutions for timelike and lightlike motion. The same holds for the doubly spinning and charged five dimensional black ring considered by Grunau et al in 2013 \cite{Grunauetal2013}.  

\subsubsection{Black strings}
Grunau and Khamesra \cite{GrunauKhamesra2013} considered particle motion in (rotating) black string space-times, which are given as generalizations of Schwarzschild and Kerr space-times including a compact fifth dimension. They derived analytical solutions for timelike and lightlike motion in both the non-rotating and the rotating space-time.

\subsection{Particle motion in cosmic string space-times}

\subsubsection{Schwarzschild black hole with cosmic string}

Owing to the possible connection to string theory, cosmic strings have gained a lot of renewed interest over the past years. They are topological defects that could have formed in one of the numerous phase transitions in the early universe. Therefore it is of principle interest to consider the possibility of black holes pierced by an infinitely thin cosmic string and to work out possible observational consequences. 

This has been carried through in \cite{Hackmannetal_SchwarzschildString_PRD10}. In this paper the geodesic motion in the space-time of a Schwarzschild black hole pierced by an infinitely thin cosmic string has been solved analytically for massive bodies and for light rays, respectively. The solutions of the geodesic equation depend on the particle's energy and angular momentum, the ratio between the component of the angular momentum aligned with the axis of the string and the total angular momentum, the deficit angle of the space-time as well as the mass of the black hole. One main difference to the orbits in the Schwarzschild case is that they are not restricted to an orbital plane. The change of the deficit angle affects the radial motion significantly. 

In \cite{Hackmannetal_SchwarzschildString_PRD10} a bound of the energy per unit length of the string was found and compared with experimental tests of general relativity. From perihelion shift and light deflection a rough estimate for the string line mass density was derived, $\mu \leq 10^{160}\,{\rm kg/m}$. The eventual existence of cosmic strings will also influence the creation of gravitational waves since the deficit angle modifies the angular velocity and, thus, the temporal change of the effective quadrupole responsible for the emitted gravitational wave.

\subsubsection{Kerr black hole with cosmic string}

For the same reasons there may exist Kerr black holes pierced by a cosmic string. Hackmann et al \cite{Hackmannetal_KerrString_PRD10} determined the complete set of solutions for this case for massive and mass less particles using the standard techniques developed for the Kerr space-time. The solutions of the geodesic equation can be classified according to the particle's energy and angular momentum as well as the mass and the specific angular momentum of the black hole. 

Also the perihelion shift and the Lense-Thirring effect for bound orbits has been discussed. It was proven that the presence of a cosmic string enhances both effects \cite{Hackmannetal_KerrString_PRD10}. Comparison of these results with experimental data from the LAGEOS satellites yields an upper bound on the energy per unit length of a string piercing the earth which is approximately $10^{16}\;{\rm kg/m}$. These results also may have relevance to the recently suggested explanation of the alignment of the polarization vector of quasars using remnants of cosmic string decay in the form of primordial magnetic field loops.

\subsection{Particle motion in further space-times}

\subsubsection{Thin-shell wormhole}
Diemer and Smolarek \cite{DiemerSmolarek2013} recently considered timelike and lightlike motion in traversable Schwarzschild and Kerr thin-shell wormholes. 

\subsubsection{Regular black holes}

Regular black holes without curvature singularity are considered as an alternative to singular black hole solution to avoid the conceptual problems related to the occurrence of a singularity. First regular black holes models were constructed by Bardeen in 1968 \cite{Bardeen1968} and later provided with a physical reasonable source by Ay\'{o}n-Beato and Garc\'{\i}a \cite{Ayon-BeatoGarcia2000}. Ay\'{o}n-Beato and Garc\'{\i}a also derived an exact regular solution of the Einstein equations coupled to a nonlinear electrodynamics, which may be called the Ay\'{o}n-Beato Garc\'{\i}a space-time \cite{AyonBeatoGarcia98}.

%

Due to the occurrence of a square root the geodesic equation in this space-time is on the first glance not of the type considered here. However, a substitution and the introduction of a new affine parameter analogously to the Mino time give a differential equation of the type \eqref{genhyperdiff} which can be solved completely \cite{Garciaetal13}. Also the perihelion shift has been calculated and its difference from the corresponding effect in a Reissner-Nordstr\"om space-time has been isolated. 

Though the orbit for a neutral particle could be solved analytically, the orbit for a charged particle leads to a mathematically different equation of motion. One obtains
\begin{equation}
y^4 + P_n(r) y^2 + P_m(r) = 0 \qquad \text{with} \qquad y := f(u) \frac{du}{d\varphi} \, ,
\end{equation}
where $P_{n,m}$ are polynomials, $u$ is a non-rational function of the radial coordinate (the substitution used before), $\varphi$ is a space-time coordinate, and $f$ is a rational function of $u$. This equation corresponds to a more general differential equation, which may be called a quartic problem according to the corresponding quartic algebraic curve. An analytical solution has not yet been found. 

\subsubsection{Ho\v{r}ava-Lifshitz gravity}

Ho\v{r}ava-Lifshitz gravity arise from quantum gravity inspired correction to General Relativity. Various black holes solutions have been studied where the metric is given by 
\begin{equation}
ds^2 = N^2(r) dt^2 - \frac{1}{f(r)} dr^2 - r^2 \left(d\vartheta^2 + \sin^2\vartheta d\varphi^2\right)
\end{equation}
with $N^2 = f = 1 + c_1 r^2 - \sqrt{c_2 r^4 + c_3 r}$ for some constants $c_1,c_2,c_3$. In the special case $c_1=-\Lambda_W$, $c_2 = 0$, and $c_3=\alpha^2\sqrt{-\Lambda_W}$, where $\Lambda_W$ is proportional to the negative cosmological constant and $\alpha \geq \frac43^{\frac34}$ is an arbitrary parameter \cite{Lueetal2009}, the geodesic equation can be reduced to the type \eqref{genellipticdiff} for lightlike motion and to the type \eqref{genhyperdiff} for timelike motion using the substitution $u = \sqrt{r}$. The analytical solution for this case was derived by Enolskii et al \cite{EnolskietalJMP12}. For $c_2 \neq 0$ we again encounter a quartic problem. 


%
%

\section{Outlook}

In this review we presented the mathematical tools needed to obtain analytical solutions of the geodesic equation in a variety of space-times. In a first step, it is possible to solve the geodesic equation in Schwarzschild, Reissner-Nordstr\"om, Taub-NUT space-times with elliptic functions. Owing to the development of the explicit analytic integration of ordinary differential equations it was possible to solve the geodesic equation also for space-times with cosmological constant. This new method can also be applied to further space-times like higher dimensional models, space-times related to nonlinear electrodynamics, or to quantum gravity inspired modifications like Ho\v{r}ava-Lifshitz gravity. Also particle motion in Gauss-Bonnet gravity models might be treated in this way. This already shows how powerful these mathematical tools are. 

However, there are also many other other phenomena and situations where these methods could be applied. As an example, the analytical solutions for geodesic motion may be extended to (i) the analytic integration of the Jacobi equation \cite{Bazanski89,BazanskiJaranowski89} with the application to, e.g., pendulum orbits, cartwheel orbits and helical orbits which may have application in satellite geodesy, (ii) the equation of motion for particles with spin in the case that the particle moves in the equatorial plane with the spin aligned perpendicular to that plane, (iii) more general for certain motions of test particles with mass multipoles. On the level of astrophysical application an analytical timing formula for the specific situation of a pulsar orbiting a supermassive black hole may be derived. Here one has to combine the analytic description of the motion of a pulsar (treated as test particle) around a massive black hole with the analytic solution for the light ray from the pulsar to the observer. This methods may also be applied for the timing between two satellites or between a satellite and a clock on Earth. On the long term perspective a further issue is to analytically solve field equations in the field of black holes. This has many applications like superradiance and quasinormal modes.


\begin{theacknowledgments}
We would like to thank A. Garcia, D. Giulini, N. G\"urlebeck, V. Kagramanova, J. Kunz, A. Macias, V. Perlick, and D. Puetzfeld for discussions. C.L. thanks A. Macias for the invitation and the hospitality. We also would like to acknowledge financial support from the German Research Foundation DFG within the Research Training Group 1620 ``Models of Gravity''.
\end{theacknowledgments}



\bibliographystyle{aipproc}   



\end{document}